**LETTER TO THE EDITOR**

# The persistence length of two-dimensional self-avoiding random walks

**E Eisenberg**[1] **and A Baram**[2]

[1] Physics Department, Princeton University, Princeton, NJ, 08544 USA
[2] Soreq NRC, Yavne 81800, Israel

E-mail: baram@soreq.gov.il

**Abstract**
The decay of directional correlations in self-avoiding random walks on the square lattice is investigated. Analysis of exact enumerations and Monte Carlo data suggest that the correlation between the directions of the first step and the $j$th step of the walk decays faster than $j^{-1}$, indicating that the persistence length of the walk is finite.

PACS numbers: 05.40.Fb, 05.10.Ln, 05.50.+q

The main characteristic of a self-avoiding random walk (SAW) is the infinite memory of the walk, resulting from the excluded volume constraint. This property raises the question whether an initial bias persists along the entire walk, and affects the distribution of the end points. This problem was first considered by Grassberger [1], who investigated the dependence of the persistence length on the number of steps of two-dimensional SAWs. The persistence length of an SAW $\langle x_n \rangle$, is defined as the average displacement of the end point, of an $n$-step walk, along the direction of the first step. Without loss of generality, we fix the origin at the starting point of the SAW, and choose the first step to be oriented in the positive $x$ direction. Based on exact enumerations of SAWs on several two-dimensional lattices, Grassberger [1] studied the $n$ dependence of the persistence length, and concluded that the persistence length diverges like $\langle x_n \rangle \approx n^w$, with $w = 0.063(10)$ (the figure in parentheses indicates the uncertainty of the last digit). On the other hand, Redner and Privman [2] suggested, based on exact enumeration and Monte Carlo (MC) data, that the divergence is logarithmic in $n$. Later scanning MC studies [3–5] found that the persistence length data could be fitted equally well by a power law and by a logarithmic function.

In this letter we revisit this problem, and show that the persistence length of a two-dimensional SAW does not diverge, neither as a power law nor logarithmically, but rather converges to a constant.

It is very difficult to determine numerically whether a quantity converges to a constant or diverges logarithmically. Therefore in this letter we prefer to investigate the persistency problem by analysing the decay rate of the angular correlation function $c_{1,j}(n)$ between the directions of the first step and the $j$th step (monomer) of a $n$-step SAW on the square lattice.

$$c_{1,j}(n) = \frac{\sum_{k=1}^{c_n} \cos(\vartheta_{1,j}(k))}{c_n} \qquad (1)$$

the index $k$ in equation (1) runs over the $c_n$ different $n$-step SAWs, and $\theta_{1,j}(k)$ is the angle between the directions of the first step and the $j$th step of that walk. The possible values of the cosines are determined by the symmetry of the lattice. On the square lattice their possible values are: 1 (parallel), 0 (perpendicular) and –1 (anti parallel). Thus $c_{1,j}(n)$ describes the average $x$ component of the $j$th step of the walk, and the persistence length is given by summation over $j$ of the angular correlations [6]:

$$\langle x_n \rangle = \sum_{j=1}^{n} c_{1,j}(n). \qquad (2)$$

For a pure random walk $c_{1,j}(n) = \delta_{1,j}$ for all $n$, and $\langle x_n \rangle = 1$. For a random walk with correlations between successive steps only, $c_{1,j}(n)$ decreases exponentially with $j$ independent of $n$, and $\langle x_n \rangle$ converges to a constant. For a SAW one expects $c_{1,j}(n) \approx j^{-\delta}$ for $n \gg j$. The asymptotic behaviour of $\langle x_n \rangle$ depends on the magnitude of the power $\delta$, for $\delta > 1$ the persistence length converges to a constant, for $\delta = 1$ it diverges logarithmically, while for $\delta < 1$ it diverges like a power law $w = 1 - \delta$. Note that the decrease in $c_{1,j}(n)$ is much faster than that of angular correlation between two steps in the interior of a long SAW, which decays like: $c_{i,j}(n) \approx (j-i)^{-\frac{1}{2}}$, where $1 \ll i < j \ll n$ (see, e.g. [1])

We start by giving some arguments about the relevant walks, and their weights. Consider a walk $\{p\}$ which starts with a step along the $+x$ direction followed by a kink at the second step (i.e. the second step is in the $\pm y$ direction—going from the site (1, 0) to (1, 1) or to (1, −1)), which does not visit the site (2, 0). Its conjugate walk $\{q\}$, defined as a reflection of $\{p\}$ about the kink, is a valid SAW, which obeys for all $j > 1 \cos(\theta_{1,j}(p)) = -\cos(\theta_{1,j}(q))$. As a result the net contribution to $c_{1,j}(n)$ from these walks vanishes. A non-vanishing contribution from walks with a kink at (1, 0) is obtained only for walks that return during the walk to the vicinity of the kink, by visiting the site (2, 0). The fraction of SAWs with a kink at (1, 0) is approximately 0.652, and their return probability to (2, 0) decays like $k^{-59/32}$ [7, 8], where $k$ is the length of the returning loop. If the (2, 0) site is visited early in the walk, resulting in a loop whose length $k$ is small compared to $j$, and no other returns to the origin occur, then, in most cases, the contribution of this walk to the persistence is cancelled by another walk, which is its reflection about the first $y$ direction step after visiting (2, 0).

Similar arguments apply to walks with a first kink at $(m, 0)$. Walks with a kink at (2, 0), for instance, contribute to $c_{1,j}(n)$ ($j > 2$) only if they return during the walk to the site (3, 0) or to the site (4, 0). The fraction of walks with a first kink at $(m, 0)$ decreases like $2/3^m$, while the number of the asymmetric sites increases like $m$. Therefore their contribution to $c_{1,j}(n)$ cannot exceed the contribution of the $m = 1$ walks.

These considerations are utilized in the following numerical procedures.

We enumerated on the square lattice all the angular correlations $c_{1,j}(n)$ up to $n = 30$. In addition we extended the range of $n$ by computing $c_{1,j}(n)$ using the pivot algorithm version of the MC method [9], for $n = 50, 70, 100, 200$. The MC statistics scanned $O(10^{11})$ iterations, which were grouped into $\sim$10 K groups of $\sim$10 M realizations each, for the purpose of calculating the statistical error. For $n = 200$ the absolute error in the MC data is about 5.0d-6, for $j = 40$ the relative error is $\sim$0.1%, while for $j = 100$ it is $\sim$0.3%.

Figure 1 shows $c_{1,j}(n)$ as a function of $1/n$ for representative $j$ values: $j = 9, 11, 13, 19$. Up to $n = 30$ the data obtained by enumerations, and for higher $n$ values by the MC method.

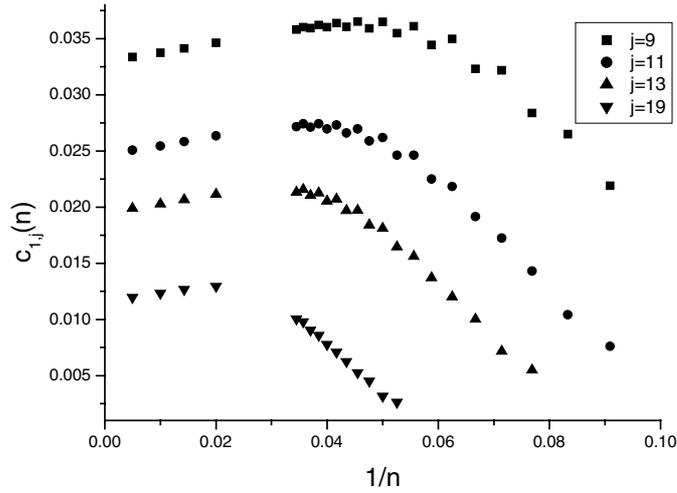

**Figure 1.** Angular correlation functions $c_{1,j}(n)$ as a function of $1/n$, for $j = 9, 11, 13, 19$.

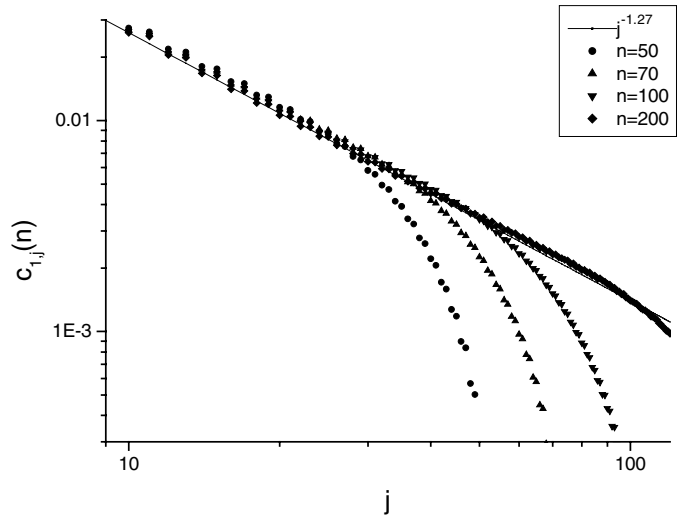

**Figure 2.** Angular correlation functions $c_{1,j}(n)$ as a function of $j$, for $n = 50, 70, 100, 200$.

For a given $j$ the angular correlation function increases monotonically with $n$ (disregard odd even effects) up to $n \sim 2j$. Then it converges linearly in $1/n$, with a positive slope, towards its asymptotic value $c_{1,j}$. As a result of the abnormally long finite size effect, it is very difficult to deduce meaningful conclusions about the asymptotic region from the enumeration data.

Figure 2 shows the MC data as a function of $j$. The main feature is the dramatic collapse of the $c_{1,j}(n)$ values for $j > n/2$. For lower $j$ values the MC data fit the $j^{-1.27}$ form quite well. We estimated the asymptotic values $c_{1,j}$ by two linear extrapolations. The first extrapolation uses the $n = 100$ and $n = 200$ MC data to obtain, up to $j = 40$, the asymptotic set $c_{1,j}(I)$:

$$c_{1,j}(I) = \tfrac{1}{100}[200 c_{1,j}(200) - 100 c_{1,j}(100)] \tag{3}$$

(for $j > 40$ $c_{1,j}(100)$ is out of the asymptotic region). A second estimation of the asymptotic values $c_{1,j}(II)$, up to $j = 28$, obtained from the $n = 200$ and $n = 70$ MC data. To reduce odd even effects we define a modified correlation function $c'_{1,j}$, given by

$$c'_{1,j} = \tfrac{1}{4}[c_{1,j-1} + c_{1,j+1} + 2c_{1,j}] \tag{4}$$

We best fitted independently the two sets and find the expressions:

$$c'_{1,j}(I) = 0.589 j^{-1.342} \tag{4a}$$

$$c'_{1,j}(II) = 0.608 j^{-1.354}. \tag{4b}$$

On the basis of these results a conservative estimation of the exponent is

$$\delta = 1.34(5). \tag{5}$$

Thus we conclude that the persistence length of a two-dimensional SAW converges to a constant.

The number of $n$-step SAWs on a lattice is $c_n \approx \mu^n n^{\gamma-1}$ where $\mu$ is the effective coordination number of the walk (connective constant), and the critical exponent $\gamma$ is a lattice independent critical exponent. For two-dimensional SAWs the value $43/32 = 1.34375$ was obtained by Nienhuis [8] for the exponent. This exact value of $\gamma$ is surprisingly similar to our estimate of $\delta$ in equation (5), but we do not find any argument to support this connection.

Finally, we note that the persistence length of a two-dimensional random walk which starts at the origin and has an excluded site (trap) at $(-1, 0)$, diverges logarithmically in $n$ [10]. The divergence results from the fact that the probability to return to the origin after $j$ steps decreases as $1/j$, and whenever such event occurs the $+x$ direction is preferred. It seems that this problem is analogous to the persistency problem for SAW, and one would expect a similar divergence of the persistence length. However, as explained above, the SAW probability to return to the origin is substantially smaller. As a result the bias at the origin affects a smaller fraction of the walks, and it does not reflect itself in a divergence of the persistence length.